\documentclass[12pt]{article}  
\usepackage{amssymb,latexsym}  
\usepackage{amsmath}  
 
\columnwidth\textwidth

\begin{document}

\newtheorem{df}{Definition}  
\newtheorem{thm}{Theorem}  
\newtheorem{lem}{Lemma}  
 
\begin{titlepage}  
 
\noindent  
 
\begin{center}  
{\LARGE Quantum model of classical mechanics:\\ maximum entropy packets}  
\vspace{1cm}

P. H\'{a}j\'{\i}\v{c}ek \\  
Institute for Theoretical Physics \\  
University of Bern \\  
Sidlerstrasse 5, CH-3012 Bern, Switzerland \\  
hajicek@itp.unibe.ch

\vspace{1cm}  
 
 
April 2009 \\  
 
 
\vspace*{5mm}  
 
\nopagebreak[4]  
 
\begin{abstract}
In a previous paper, a statistical method of constructing quantum models of
classical properties has been described. The present paper concludes the
description by turning to classical mechanics. The quantum states that
maximize entropy for given averages and variances of coordinates and momenta
are called ME packets. They generalize the Gaussian wave packets. A
non-trivial extension of the partition-function method of probability calculus
to quantum mechanics is given. Non-commutativity of quantum variables limits
its usefulness. Still, the general form of the state operators of ME packets
is obtained with its help. The diagonal representation of the operators is
found. A general way of calculating averages that can replace the partition
function method is described. Classical mechanics is reinterpreted as a
statistical theory. Classical trajectories are replaced by classical ME
packets. Quantum states approximate classical ones if the product of the
coordinate and momentum variances is much larger than Planck constant. Thus,
ME packets with large variances follow their classical counterparts better
than Gaussian wave packets. 
\end{abstract}

\end{center}

\end{titlepage}

\section{Introduction}
'The quantum origin of the classical' \cite{Zurek} is a non trivial open
problem of quantum theory: 'how to explain within quantum theory the classical
appearance of our macroscopic world' \cite{Zeh}. The purpose of the present
paper is to explain the classical properties as specific properties of quantum
systems. The conceptual structure of quantum mechanics and with it the
foundation of modern physics cannot be completely understood without such
explanation. 

Every existing attempt in this direction starts with the assumption that the
basic properties of individual quantum systems are single values of
observables and that all other properties can be constructed or derived from
these basic ones. Then, quantum mechanics does not admit any genuine realist
interpretation. The definitive account is given by the Bub-Clifton-Goldstein
theorem \cite{Bub}. Only different kinds of apparent realism for different
restricted sets of properties are possible. It is then difficult to explain
how objective classical properties can emerge within quantum mechanics. 

In \cite{PHJT}, we have initiated a very different approach. Our two main
starting points are: 
\begin{enumerate}
\item Value of an observable $o$ of an individual quantum system $A$ is not a
  property of $A$ alone, but of a composed system $A + M$, where $M$ is an
  apparatus measuring $o$. The value of $o$ measured by $M$ is not determined
  (in general) before its measurement by $M$. It is created by the measurement
  process. We call, therefore, single values of observables measurable on $A$
  {\em extrinsic properties} of $A$. The whole existing quantum mechanics is
  practically only the theory of the extrinsic properties. 
\item As properties of a quantum system $A$, we allow also quantities that a)
  have values that may be more complex mathematical objects than just real
  numbers (such as sets, mappings between sets, etc.) and b) such a value need
  not be directly observable in a single measurement. 
\end{enumerate}
Then, there are properties of quantum systems that can be viewed as
determinate before their measurements without any other condition. They have
been called {\em intrinsic}, listed and classified into {\em structural} and
{\em conditional} in \cite{PHJT}. Structural are those that are uniquely
determined by the kind of quantum system (systems of the same kind are
indistinguishable in the well-known strong and exclusively quantum-mechanical
sense). Conditional are those that are uniquely determined by preparations. In
\cite{PHJT}, a new realist interpretation of quantum mechanics has been
described based on the intrinsic properties. 

Some of the intrinsic properties have been proposed as quantum models of
classical properties in \cite{PHJT}. In particular, two kinds of conditional
properties have been important: averages of quantum observables (including
their variances) in a prepared state and the von-Neumann entropy of the
state. Some explanatory remarks may be helpful. First, any preparation is
defined by physical (objective) conditions. It need not be a process carried
out by humans. Second, in the approach of \cite{PHJT}, an average of an
observable is not constructed or derived from the ''more basic'' single values
of the observable. Averages are determined uniquely by preparation, single
values are not. Not the single values determine the average but the average
restrict possible single values. 

Third, entropy is often considered as a measure of observer ignorance and,
{\em therefore}, as a subjective concept. The ''therefore'' is
fallacious. Entropy can generally be defined as follows. Let $\Gamma$ be a
complete set of mutually exclusive properties or states of a system $A$ that
is a measurable set and let the measure be $\mu(\rho)$ for
$\rho\in\Gamma$. Let physical condition $C$ on $A$ lead to restriction on
possible properties or states of $A$ such that the probability of the state
$\rho$ to occur under $C$ is $p(\rho)$. Then the entropy $S(C) = -\int_\Gamma
\mu(\rho)p(\rho)$. 

Such entropy can be a measure of ignorance in the following sense: everything
we may know on $A$ are some physical properties that therefore define some
physical condition $C$. Then the above value of entropy gives the objective
uncertainty on the properties or states of the system associated with
condition $C$. Thus it is, in this case, if we know only $C$, simultaneously a
measure of our ignorance on the system. 

Let us now briefly review the most popular approaches to the problem of
classicality. At the present time, the problem does not seem to be solved in a
satisfactory way, the shortcoming of the approaches being well known
\cite{d'Espagnat, bell2,Wallace}. We mention them only fleetingly. First, the
quantum decoherence theory \cite{Zurek, Zeh} works only if certain observables
concerning both the environment and the quantum system cannot be measured (see
the analysis in \cite{d'Espagnat,Bub}). The deep reason is that one works with
values of observables. Second, the theories based on coarse-grained operators
\cite{Peres,poulin,Kofler}: the problem is the same as with the
decoherence. For example, the Legget-Garg inequality \cite{Kofler} is a
condition for the validity of the principle of macroscopic realism that works
with values of observables. Third, the Coleman-Hepp theory
\cite{Hepp,Bell3,Bona} and its modifications \cite{Sewell,Primas}: they are
based on some particular theorems that hold for infinite systems but do not
hold even approximately for finite ones (see the analysis in \cite{Bell3}). 

The approach of \cite{PHJT} is free of these shortcomings.First, intrinsic
properties are quantum properties of all quantum systems and there is no
question about how they emerge in quantum mechanics. This avoids e.g. the
artificial construction of classical propeties in the Coleman-Hepp
approach. Second, they are considered as, and proved to be, objective in
\cite{PHJT}. Hence, second, they could in principle serve as classical
properties because they can satisfy the principle of classical realism. This
avoids the problems of both the quantum-decoherence and the coarse-grained
theory that assume values of quantum observables to be basic properties. 

More specifically, \cite{PHJT} has conjectured that intrinsic averages and
entropy for certain macroscopic quantum systems can model all their classical
properties. Classical states of a macroscopic quantum system $T$ have been
defined as determined by averages $O_1,\cdots,O_k$ of quantum operators
$o_1,\cdots,o_k$ that form a small subset of the algebra of observables of
$T$. Finally, our modelling or construction of classical properties is nothing
but statistical physics. The statistical methods that were highlighted in
\cite{PHJT} can only work if the following hypothesis (basic hypothesis of
statistical physics) is correct: The overwhelming part of macroscopic systems
occur in quantum states that maximize entropy under the conditions of given
averages $O_1,\cdots,O_k$. This hypothesis is supported by observation and can
be derived from quantum mechanics for a class of simplified models such as
\cite{Gemmer,Linden,Lebowitz}. 

The problem of quantum measurement (see, e.g.,
\cite{d'Espagnat,Isham,Peres,Wallace} and references listed there) can be
formulated within our approach as follows. For a measurement by a quantum
apparatus $M$ on a quantum system $A$, there must be an interaction between
$M$ and $A$ as well as processes in $M$ satisfying the conditions: (a) $M$
changes its classical state as the result of the interaction, (b) the change
depends on the initial quantum state of $A$, (c) average values defining
different resulting classical states must differ by much more than the values
of their variances, and (d) Born rule is fulfilled. The knowledge of what
properties of $M$ can be considered as classical is the first step. Still, to
construct a model of interaction and processes in $M$ satisfying conditions
(a), (b), (c) and (d) remains a non trivial problem. Ref. \cite{PHJT} and the
present paper do not offer a solution to the problem of measurement. 

A way of model construction for internal (thermodynamic) properties of
macroscopic quantum systems was described in detail in \cite{PHJT}. This,
however, did not work for external (mechanical) properties of such
systems. The aim of the present paper is to fill in this gap. 

The plan of the paper is as follows. In Sec.\ 2, classical mechanics is
interpreted as a statistical theory. The existence of sharp trajectories is
rejected so that all possible states of systems are described by fuzzy
distribution functions. One choice for such distributions are the so-called
maximum-entropy packets (ME-packet). These are states that maximize entropy
for given averages and variances of coordinates and momenta. The method of
partition function is used to calculate the general form of the distribution
function. For a simple solvable example, the dynamical equations for the
averages and variances are obtained. The example shows how the equations of
motion are reinterpreted in our theory. For general potentials, we use an
approximative method: step-by-step calculation of the higher and higher time
derivatives of coordinates and momenta. This will later be compared with
quantum ME packets. 

Turning to quantum mechanics, we apply the maximum-entropy principle in an
analogous way in Sec.\ 3. The averages and variances are taken over from the
classical states that are to be modelled. A straightforward generalization of
the partition function method is now complicated by the non-commutativity of
the coordinates and momenta. We can show that only the first derivatives of
the logarithm of partition function have the usual meaning. This is, however,
sufficient for calculating the state operators for all ME packets. We find the
diagonal representation of the state operator in Sec.\ 3.2 and obtain with its
help the general form of the partition function and the state operator
itself. It turns out that Gaussian wave packets are special case of ME
packets, namely those with zero entropy and minimal uncertainty. The diagonal
representation gives us also a powerful method to calculate averages of higher
moments. In fact, what has been done in Secs.\ 3.1 and 3.2 is a non-trivial
extension of the partition-function method of the probability calculus as
described, e.g., in \cite{Jaynes} to quantum mechanics that might also be of
some interest for mathematicians. 

In Sec.\ 3.3, the equations of motion are calculated in analogy to the
classical case. We find that the quantum corrections to the classical
equations come only from high powers of $q$ in the expansion of the potential
or in high powers of $t$ in the expansion of the time-dependent
averages. Also, these corrections are of the second order in $\hbar$. These
results show that our quantum models follows classical trajectories very
closely. The nature of classical limit is studied in Sec.\ 4. The result,
which may seem surprising, is that it is the limit of large variances, not
small. Thus, quantum ME packets with large variances follow their classical
counterparts better than Gaussian wave packets. Of course, the way we measure
the size of the variances is important here. The variances that are large with
respect to this measure can still be sufficiently small to agree with
observations. Finally, Sec.\ 5 concludes the paper by summarizing the main
ideas and the most important results.

\section{Statistical form of classical mechanics}
Let us start with the warning that the topic of this section has nothing to do
with what is usually called 'statistical mechanics'. 

If one is going to model classical mechanics then what are the properties that
one would like to reproduce? The most conspicuous property from the point of
view of quantum mechanics appears to be the sharpness of mechanical
trajectories in the phase space because quantum mechanics denies the existence
of such trajectories. This leads most researchers to aim at quantum states the
phase-space picture of which is as sharp as possible. That are states with
minimum uncertainty allowed by quantum mechanics. For one degree of freedom,
described by coordinate $q$ and momentum $p$, the uncertainty is given by the
quantity 
\begin{equation}\label{uncertainty}
\nu = \frac{2\Delta q\Delta p}{\hbar},
\end{equation}where $\Delta a$ is the variance of quantity $a$,
\begin{equation}\label{variance}
\Delta a = \sqrt{\langle a^2\rangle -\langle a\rangle^2}.
\end{equation}
It is well known that minimum uncertainty allowed by quantum mechanics is $\nu
= 1$.  

The states with $\nu = 1$ are, however, very special states. First, they must
be pure states such as Gaussian wave packets or coherent states. Such states
are very difficult to prepare unlike the usual states of macroscopic systems
described by classical mechanics. They are also prone to strong distortion by
measurements. Moreover, as pure states, they can be linearly superposed. This
is another peculiarity that is never observed for states of systems of
classical mechanics. Hence, trying to get a trajectory as sharp as possible
leads to the loss of other desirable properties. 

Moreover, observations within classical mechanics admit the notion that the
sharpness of phase-space trajectories is only a mathematical and methodical
feature of classical mechanics. It may be just an idealisation, a limit in
which things become mathematically simpler. We can use it in calculations
which, however, must also take into account the necessary non-zero variances
of real observations. Indeed, such observations are generally afflicted with
uncertainties $\nu \gg 1$. Hence, if we want to compare the predictions of our
quantum models with observations of classical mechanics, we are forced to
compare states that are fuzzy in both theories. 

One idea of the present paper is to consider states with given averages and
variances of the coordinates and momenta and leave everything else as fuzzy as
possible. To calculate the corresponding probability distributions in
classical, and the state operators in quantum mechanics, we shall, therefore,
apply the maximum entropy principle. This is a general principle in
mathematical theory of probabilities (see \cite{Jaynes}) and it should not be
confused with the well-known thermodynamic law. The resulting states are
called {\em maximum-entropy packets}, ME-packets. The averages of coordinates
and momenta take over the role of coordinate and momenta in classical
mechanics. In any case the averages represent measurable aspects of these
variables. The dynamical evolution of variances is an important indicator of
the applicability of the model one is working with. It determines the time
intervals within which reasonable predictions are possible. 

Consider a three-body system that is to model the Sun, Earth and Jupiter. It
turns out that generic trajectories starting as near to each other as, say,
the dimension of the irregularities of the Earth surface will diverge from
each other by dimensions of the Earth-Sun distance after the time of only
about ten million years. This seems to contradict the four billion years of
relatively stable Earth motion around the Sun that is born out by
observations. The only way out is the existence of a few special trajectories
that are much stabler than the generic ones and the fact that bodies following
an unstable trajectory have long ago fallen into the Sun or have been ejected
from the solar system. 

An important question is that on the ontological status of ME-packets and on
the nature of the limit in which trajectories become sharp. The usual
standpoint is that any mechanical system always objectively {\em is} in a
state of a completely sharp trajectory. Any more fuzzy state is only the
result of our incomplete knowledge. Thus, the fuzzy states are not considered
themselves as real. Here, we take the opposite standpoint. For us, a state to
be real, it must be determined by objective initial conditions. A simple
example is a gun in a position that is fixed in a reproducible way and that
shoot bullets using cartridges of a given provenance. The state of each
individual shot is defined by the conditions and is the same for all shots
even if observations may have different results for different shots. A finer
analysis is possible only as long as new initial conditions are specified that
determine a subset of individual shots. In the theoretical description of a
state, we can make the limit of $\Delta Q \rightarrow 0,\Delta P \rightarrow
0$. This is considered as a non-existing, but practically useful
idealization. 

To limit ourselves just to given averages and variances of coordinates and
momenta is a great simplification that enables us to obtain interesting
results easily. Some further discussion on quantum modelling of classical
properties is in order. On the one hand, for internal degrees of freedom, the
usual thermodynamic methods give small relative variances as a consequence of
the state coordinates being extensive, the entropy being maximal and the
system being macroscopic. This does not work for external (mechanical)
properties. The difference is due to the simple fact that the internal degrees
of freedom are not accessible to manipulation and have small variances
spontaneously. The external degrees of freedom are accessible to manipulations
and it is easy to prepare states with small as well as large variances. There
is no objective need that the variances are small spontaneously. The idea that
really existing mechanical states must always have only small variances is
caused by a purely theoretical notion that all real mechanical systems have an
absolutely sharp phase space trajectory and this notion is clearly false. The
only problem is that it has become a part of our subconscious psychology. 

On the other hand, if quantum ME packets are to be quantum models of the
classical ME packets with the same averages and variances then this is a more
general situation than that considered in Ref.\ \cite{PHJT}. Quantum ME packet
is a classical state in the sense of Ref.\ \cite{PHJT} if it has small
variances. Only then, the average values are directly observable on individual
systems.

\subsection{Classical ME-packets}
Let us first consider systems with one degree of freedom. The generalization
to any number is easy. Let the coordinate be $q$ and the momentum $p$. A state
is a distribution function $\rho(q,p)$ on the phase space spanned by $q$ and
$p$. The function $\rho(q,p)$ is dimension-free and normalized by 
$$
\int\frac{dq\,dp}{v}\,\rho = 1\ ,
$$
where $v$ is an auxiliary phase-space volume to make the integration
dimension-free. The entropy of $\rho(q,p)$ can be defined by 
$$
S := -\int\frac{dq\,dp}{v}\,\rho \ln\rho\ .
$$
The value of entropy will depend on $v$ but the most of other results will
not. Classical mechanics does not offer any idea of how to fix $v$. We shall
get its value from quantum mechanics. 

Let us define: ME-packet is the distribution function $\rho$ that maximizes
the entropy subjected to the conditions: 
\begin{equation}\label{21.4}
\langle q\rangle = Q\ ,\quad \langle q^2\rangle = \Delta Q^2 + Q^2\ ,
\end{equation}
and
\begin{equation}\label{21.5}
\langle p\rangle = P\ ,\quad \langle p^2\rangle = \Delta P^2 + P^2\ ,
\end{equation}
where $Q$, $P$, $\Delta Q$ and $\Delta P$ are given values of averages and
variances of $q$ and $p$. We have used the abbreviation 
$$
\langle x\rangle = \int\frac{dq\,dp}{v}\,x\ .
$$

The explicit form of $\rho$ can be found using the partition-function method
as described e.g.\ in \cite{Jaynes}. The variational principle yields 
\begin{equation}\label{rho}
\rho = \frac{1}{Z(\lambda_1,\lambda_2,\lambda_3,\lambda_4)}\exp(-\lambda_1 q
-\lambda_2 p -\lambda_3 q^2 - \lambda_4 p^2)\ , 
\end{equation}
where
$$
Z = \int\,\frac{dq\,dp}{v}\exp(-\lambda_1 q -\lambda_2 p -\lambda_3 q^2 -
\lambda_4 p^2)\ , 
$$
and $\lambda_1$, $\lambda_2$, $\lambda_3$ and $\lambda_4$ are the Lagrange
multipliers. Hence, the partition function for classical ME-packets is given
by 
\begin{equation}\label{22.1}
Z= \frac{\pi}{v}\frac{1}{\sqrt{\lambda_3\lambda_4}}
\exp\left(\frac{\lambda_1^2}{4\lambda_3} +
  \frac{\lambda_2^2}{4\lambda_4}\right)\ . 
\end{equation}
The expressions for $\lambda_1$, $\lambda_2$, $\lambda_3$ and $\lambda_4$ in
terms of $Q$, $P$, $\Delta Q$ and $\Delta P$ can be obtained by solving the
equations 
$$
\frac{\partial\ln Z}{\partial\lambda_1} = -Q\ , \quad \frac{\partial\ln
  Z}{\partial\lambda_3} = -\Delta Q^2 - Q^2\ , 
$$
and
$$
\frac{\partial\ln Z}{\partial\lambda_2} = -P\ , \quad \frac{\partial\ln
  Z}{\partial\lambda_4} = -\Delta P^2 - P^2\ . 
$$
The result is:
\begin{equation}\label{22.2}
\lambda_1 = -\frac{Q}{\Delta Q^2}\ ,\quad \lambda_3 = \frac{1}{2\Delta Q^2}\ ,
\end{equation}
and
\begin{equation}\label{22.3}
\lambda_2 = -\frac{P}{\Delta P^2}\ ,\quad \lambda_4 = \frac{1}{2\Delta P^2}\ .
\end{equation}
Substituting this into Eq.\ (\ref{rho}), we obtain the distribution function
of a one-dimensional ME packet. The generalization to any number of dimensions
is trivial. 
\begin{thm}
The distribution function of the ME-packet for a system with given averages
and variances $Q_1,\cdots,Q_n$, $\Delta Q_1,\cdots,\Delta Q_n$ of coordinates
and $P_1,\cdots,P_n$, $\Delta P_1,\cdots,\Delta P_n$ of momenta, is 
\begin{equation}\label{23.1}
\rho = \left(\frac{v}{2\pi}\right)^n\prod_{k=1}^n\left(\frac{1}{\Delta
    Q_k\Delta P_k}\exp\left[-\frac{(q_k-Q_k)^2}{2\Delta Q_k^2}
    -\frac{(p_k-P_k)^2}{2\Delta P_k^2}\right]\right)\ . 
\end{equation}
\end{thm}
We observe that all averages obtained from $\rho$ are independent of $v$ and
that the result is a Gaussian distribution in agreement with Jaynes'
conjecture that the maximum entropy principle gives the Gaussian distribution
if the only conditions are fixed values of the first two moments. 

As $\Delta Q$ and $\Delta P$ approach zero, $\rho$ becomes a delta-function
and the state becomes sharp. For some quantities, this limit is sensible for
others it is not. In particular, the entropy, which can easily be calculated, 
$$
S = 1 + \ln\frac{2\pi\Delta Q\Delta P}{v}\ ,
$$
diverges to $-\infty$. This is due to a general difficulty in giving a
definition of entropy for a continuous system that would be satisfactory in
every respect. What one could do is to divide the phase space into cells of
volume $v$ so that $\Delta Q\Delta P$ could not be chosen smaller than
$v$. Then, the limit $\Delta Q\Delta P \rightarrow v$ of entropy would make
more sense. 

The average of any monomial of the form $q^k p^l q^{2m} p^{2n}$ can be
calculated with the help of partition-function method as follows: 
\begin{equation}\label{43.1}
\langle q^k p^l q^{2m} p^{2n}\rangle = \frac{(-1)^{\mathbf N}}{Z}\ 
\frac{\partial^{\mathbf N} Z}{\partial\lambda_1^k \partial\lambda_2^l
  \partial\lambda_3^m \partial\lambda_4^n }\ , 
\end{equation}
where ${\mathbf N} = k+l+2m+2n$, $Z$ is given by Eq. (\ref{22.1}) and the
values (\ref{22.2}) and (\ref{22.3}) must be substituted for the Lagrange
multipliers after the derivatives are taken. 

Observe that this enables to calculate the average of a monomial in several
different ways. Each of these ways, however, leads to the same result due the
the identities 
$$
\frac{\partial^2 Z}{\partial \lambda_1^2} = -\frac{\partial Z}{\partial
  \lambda_3}\ ,\quad \frac{\partial^2 Z}{\partial \lambda_2^2} =
-\frac{\partial Z}{\partial \lambda_4}\ , 
$$
which are satisfied by the partition function.

\subsection{Equations of motion}
Let us assume that the Hamiltonian of our system has the form
\begin{equation}\label{35.1}
H = \frac{p^2}{2m} + V(q)\ ,
\end{equation}
where $m$ is the mass and $V(q)$ the potential function. The equations of
motion are 
$$
\dot{q} = \{q,H\}\ ,\quad\dot{p} = \{p,H\}\ .
$$
Inserting (\ref{35.1}) for $H$, we obtain
\begin{equation}\label{36.6}
\dot{q} = \frac{p}{m}\ ,\quad\dot{p} = -\frac{dV}{dq}\ .
\end{equation}
The general solution to these equations can be written in the form 
\begin{equation}\label{36.65}
q(t) = q(t;q,p)\ ,\quad p(t) = p(t;q,p)\ ,
\end{equation}
where
\begin{equation}\label{36.3}
q(0;q,p) = q\ ,\quad p(0;q,p) = p\ ,
\end{equation}
$q$ and $p$ being arbitrary initial values. We obtain the equations of motion
for the averages and variances: 
\begin{equation}\label{36.7}
Q(t) = \langle q(t; q,p)\rangle\ ,\quad \Delta Q(t) = \sqrt{\langle (q(t;q,p)-
  Q(t))^2\rangle} 
\end{equation}
and
\begin{equation}\label{36.8}
P(t) = \langle p(t; q,p)\rangle\ ,\quad \Delta P(t) = \sqrt{\langle (p(t;q,p)-
  P(t))^2}\rangle\ . 
\end{equation}
In general, $Q(t)$ and $P(t)$ will depend not only on $Q$ and $P$, but also on
$\Delta Q$ and $\Delta P$. 

Let us consider the special case of at most quadratic potential:
\begin{equation}\label{36.1}
V(q) = V_0 + V_1 q + \frac{1}{2} V_2 q^2\ ,
\end{equation}
where $V_k$ are constants with suitable dimensions. If $V_1 = V_2 =0$, we have
a free particle, if $V_2 = 0$, it is a particle in a homogeneous force field
and if $V_2 \neq 0$, it is an harmonic or anti-harmonic oscillator. 

In this case, the general solution has the form
\begin{eqnarray}\label{37.1}
q(t) &=& f_0(t) + q f_1(t) + p f_2(t)\ , \\
\label{37.2}
p(t) &=& g_0(t) + q g_1(t) + p g_2(t)\ ,
\end{eqnarray}
where $f_0(0) = f_2(0) = g_0(0) = g_1(0) = 0$ and $f_1(0) = g_2(0) = 1$. If
$V_2 \neq 0$, the functions are 
\begin{equation}\label{37.4}
f_0(t) = -\frac{V_1}{V_2}(1-\cos\omega t)\ ,\quad f_1(t) = \cos \omega t\
,\quad f_2(t) = \frac{1}{\xi}\sin\omega t\ , 
\end{equation}
\begin{equation}\label{37.5}
g_0(t) = -\xi\frac{V_1}{V_2}\sin\omega t\ ,\quad g_1(t) = -\xi\sin \omega t\
,\quad g_2(t) = \cos\omega t\ , 
\end{equation}
where
$$
\xi = \sqrt{mV_2}\ ,\quad \omega = \sqrt{\frac{V_2}{m}}\ .
$$
Only for $V_2 > 0$, the functions remain bounded. If $V_2 = 0$, we obtain
\begin{equation}\label{37.9a}
f_0(t) = -\frac{V_1}{2m}t^2\ ,\quad f_1(t) = 1\ ,\quad f_2(t) = \frac{t}{m}\ ,
\end{equation}
\begin{equation}\label{37.9b}
g_0(t) = -V_1t\ ,\quad g_1(t) = 0\ ,\quad g_2(t) = 1\ .
\end{equation}

The equations for averages and variances resulting from Eqs.\ (\ref{36.65}),
(\ref{21.4}) and (\ref{21.5}) are 
\begin{equation}\label{38.3}
Q(t) = f_0(t) + Q f_1(t) + P f_2(t)\ ,
\end{equation}
and
\begin{multline}\label{38.4}
\Delta Q^2(t) + Q^2(t) = f_0^2(t) + (\Delta Q^2 + Q^2) f_1^2(t) + (\Delta P^2
+ P^2) f_2^2(t)\\ + 2Qf_0(t)f_1(t) + 2Pf_0(t)f_2(t) + 2\langle qp\rangle
f_1(t)f_2(t)\ . 
\end{multline}
For the last term, we have from Eq.\ (\ref{43.1})
$$
\langle qp\rangle = \frac{1}{Z}\frac{\partial^2
  Z}{\partial\lambda_1\partial\lambda_2}\ . 
$$
Using Eqs.\ (\ref{22.1}), (\ref{22.2}) and (\ref{22.3}), we obtain from Eq.\
(\ref{38.4}) 
\begin{equation}\label{39.1}
\Delta Q(t) = \sqrt{f_1^2(t)\Delta Q^2 + f_2^2(t)\Delta P^2}\ .
\end{equation}
Similarly,
\begin{eqnarray}\label{39.2}
P(t) &=& g_0(t) + Q g_1(t) + P g_2(t)\ ,\\
\label{39.3}
\Delta P(t) &=& \sqrt{f_g^2(t)\Delta Q^2 + g_2^2(t)\Delta P^2}\ .
\end{eqnarray}
We observe: if functions $f_1(t)$, $f_2(t)$, $g_1(t)$ and $g_2(t)$ remain
bounded, the variances also remain bounded and the predictions are possible in
arbitrary long intervals of time. Otherwise, there will always be only limited
time intervals in which the theory can make predictions. 

In the case of general potential, the functions (\ref{36.65}) can be expanded
in products of powers of $q$ and $p$, and the averages of these products will
contain powers of the variances. However, as one easily sees form formula
(\ref{43.1}) and (\ref{22.1}), 
$$
\langle q^kp^l\rangle = Q^kP^l +X\Delta Q +Y\Delta P\ ,
$$
where $X$ and $Y$ are bounded functions. It follows that the dynamical
equations for averages coincide, in the limit $\Delta Q \rightarrow 0, \Delta
P \rightarrow 0$, with the exact dynamical equations for $q$ and $p$. It is an
idealisation that we consider as not realistic, even in principle, but that
may still be useful for calculations. 

Let us expand a general potential function in powers of $q$,
\begin{equation}\label{50.2}
V(q) = \sum_{k=0}^\infty \frac{1}{k!}V_k q^k\ ,
\end{equation}
where $V_k$ are constants of appropriate dimensions. The Hamilton equations
can be used to calculate all time derivatives at $t=0$. First, we have 
$$
\frac{dq}{dt} = \{q,H\} = \frac{p}{m}\ .
$$
This equation can be used to calculate all derivatives of $q$ in terms of
those of $p$: 
\begin{equation}\label{50.4}
\frac{d^nq}{dt^n} = \frac{1}{m}\frac{d^{n-1}p}{dt^{n-1}}\ .
\end{equation}

A simple iterative procedure gives us further time derivatives of $p$:
\begin{equation}\label{50.5}
\frac{dp}{dt} = -V_1-V_2q-\frac{V_3}{2}q^2-\frac{V_4}{6}q^3 + r_5\ ,
\end{equation}
\begin{equation}\label{50.6}
\frac{d^2p}{dt^2} = -\frac{V_2}{m}p-\frac{V_3}{m}qp-\frac{V_4}{2m}q^2p + r_5\ ,
\end{equation}
\begin{multline}\label{50.7}
\frac{d^3p}{dt^3} = -\frac{V_3}{m^2}p^2-\frac{V_4}{m^2}qp^2 + \frac{V_1V_2}{m}
+ \frac{V_1V_3+V_2^2}{m}q + \frac{3V_2V_3+V_1V_4}{2m}q^2 \\ +
\frac{4V_2V_4+3V_3^2}{6m}q^3 + \frac{5V_3V_4}{12m}q^4 + \frac{V_4^2}{12m}q^5 +
r_5\ , 
\end{multline}
and
\begin{multline}\label{50.8}
\frac{d^4p}{dt^4} = -\frac{V_4}{m^3}p^3 + \frac{3V_1V_3+V_2^2}{m^2}p +
\frac{3V_1V_4+5V_2V_3}{m^2}qp + \frac{5V_3^2+8V_2V_4}{2m^2}q^2p \\ +
3\frac{V_3V_4}{m^2}q^3p + \frac{3V_4^2}{4m^2}q^4p + r_5\ , 
\end{multline}
where $r_k$ is the rest term that is due to all powers in (\ref{50.2}) that
are not smaller than $k$ (the rests symbolize different expressions in
different equations). The purpose of having all time derivatives up to the
fourth order is to show later that it is the highest order in which no quantum
corrections appear in the equations for the averages. 

Taking the average of both sides of Eqs.\ (\ref{50.5})--(\ref{50.8}), and
using Eq.\ (\ref{43.1}), (\ref{22.1})--(\ref{22.3}), we obtain 
\begin{equation}\label{51.1}
\frac{dP}{dt} = -V_1-V_2Q-\frac{V_3}{2}Q^2-\frac{V_4}{6}Q^3
-\frac{V_3+V_4Q}{2}\Delta Q^2 + r_5\ , 
\end{equation}
\begin{equation}\label{51.2}
\frac{d^2P}{dt^2} = -\frac{V_2}{m}P + \frac{V_3}{m}QP + \frac{V_4}{2m}Q^2P +
\frac{V_4}{2m}P\Delta Q^2 + r_5\ , 
\end{equation}
\begin{multline}\label{51.3}
\frac{d^3P}{dt^3} = -\frac{V_3}{m^2}P^2-\frac{V_4}{m^2}QP^2 + \frac{V_1V_2}{m}
+ \frac{V_1V_3+V_2^2}{m}Q + \frac{3V_2V_3+V_1V_4}{2m}Q^2 \\+
\frac{4V_2V_4+3V_3^2}{6m}Q^3 + \frac{5V_3V_4}{12m}Q^4 +
\frac{V_4^2}{12m}Q^5-\left(\frac{V_3}{m^2}+\frac{V_4}{m^2}Q\right)\Delta P^2
\\+ \left(\frac{3V_2V_3+V_1V_4}{2m} + \frac{4V_2V_4+3V_3^2}{2m}Q +
  \frac{5V_3V_4}{2m}Q^2 + \frac{5V_3V_4}{4m}\Delta Q^2 \right. \\+
\left. \frac{5V_4^2}{6m}Q^3 + \frac{5V_4^2}{4m}Q\Delta Q^2\right)\Delta Q^2 +
r_5\ , 
\end{multline}
and
\begin{multline}\label{51.4}
\frac{d^4P}{dt^4} = -\frac{V_4}{m^3}P^3 + \frac{3V_1V_3+V_2^2}{m^2}P +
\frac{3V_1V_4+5V_2V_3}{m^2}QP \\+ \frac{5V_3^2+8V_2V_4}{2m^2}Q^2P  +
3\frac{V_3V_4}{m^2}Q^3P + \frac{3V_4^2}{4m^2}Q^4P-\frac{3V_4}{m^3}P\Delta P^2
\\+\left(\frac{5V_3^2+8V_2V_4}{2m^2}P + \frac{9V_3V_4}{m^2}QP +
  \frac{9V_4^2}{2m^2}Q^2P + \frac{9V_4^2}{4m^2}P\Delta Q^2\right)\Delta Q^2 +
r_5\ . 
\end{multline}

We can see, that the limit $\Delta Q\rightarrow 0,\Delta P \rightarrow 0$ in
Eqs.\ (\ref{51.1})-(\ref{51.4}) lead to equations that coincide with Eqs.\
(\ref{50.5})-(\ref{50.8}) if $Q\rightarrow q,P\rightarrow p$ as promised.

\section{Quantum ME-packets}
Let us now turn to quantum mechanics and try to solve an analogous
problem. Let a system with one degree of freedom be described by the operators
$q$ and $p$ and let us look for a state $\rho$, a normalized, 
$$
\text{Tr}\rho = 1\ ,
$$
self-adjoint positive operator, that maximizes von Neumann entropy
\begin{equation}\label{VNE}
S = \text{Tr}(\rho\ln\rho)
\end{equation}
under the conditions
\begin{equation}\label{12.1}
\text{Tr}(\rho q) = Q\ ,\quad \text{Tr}(\rho q^2) = Q^2 + \Delta Q^2\ ,
\end{equation}
\begin{equation}\label{12.2}
\text{Tr}(\rho p) = P\ ,\quad \text{Tr}(\rho p^2) = P^2 + \Delta P^2\ ,
\end{equation}
where $Q$, $P$, $\Delta Q$ and $\Delta P$ are given numbers. The states that
satisfy these conditions are called quantum ME-packets. 

\subsection{Calculation of the state operator}
To solve the mathematical problem, we use the method of Lagrange multipliers
as in the classical case. Thus, the following equation results: 
\begin{equation}\label{12.5}
dS -\lambda_0 d\text{Tr}\rho-\lambda_1 d\text{Tr}(\rho q)-\lambda_2
d\text{Tr}(\rho p) -\lambda_3 d\text{Tr}(\rho q^2)-\lambda_4 d\text{Tr}(\rho
p^2) = 0\ . 
\end{equation}
The differentials of the terms that are linear in $\rho$ are simple to
calculate: 
$$
d\text{Tr}(\rho x) = \sum_{mn}x_{nm}d\rho_{mn}.
$$
Although not all elements of the matrix $d\rho_{mn}$ are independent (it is a
hermitian matrix), we can proceed as if they were because the matrix $x_{nm}$
is to be also hermitian. The only problem is to calculate $dS$. We have the
following 
\begin{lem}
\begin{equation}\label{10.1}
dS = -\sum_{mn}[\delta_{mn} + (\ln\rho)_{mn}]d\rho_{mn}\ .
\end{equation}
\end{lem}
{\bf Proof}
Let $M$ be a unitary matrix that diagonalizes $\rho$,
$$
M^\dagger \rho M = R\ ,
$$
where $R$ is a diagonal matrix with elements $R_n$. Then $S = -\sum_n R_n\ln
R_n$. Correction to $R_n$ if $\rho \mapsto \rho + d\rho$ can be calculated by
the first-order formula of the stationary perturbation theory. This theory is
usually applied to Hamiltonians but it holds for any perturbed hermitian
operator. Moreover, the formula is exact for infinitesimal
perturbations. Thus, 
$$
R_n \mapsto R_n + \sum_{kl}M^\dagger_{kn}M_{ln}d\rho_{kl}\ .
$$
In this way, we obtain
\begin{multline*}
dS = -\sum_n\left(R_n + \sum_{kl}M^\dagger_{kn}M_{ln}d\rho_{kl}\right)\\
\times \ln\left[R_n\left(1 +
    \frac{1}{R_n}\sum_{rs}M^\dagger_{rn}M_{sn}d\rho_{rs}\right)\right] -\sum_n
R_n\ln R_n\\ 
= -\sum_n\left[\ln R_n\sum_{kl}M^\dagger_{kn}M_{ln}d\rho_{kl}
  +\sum_{kl}M^\dagger_{kn}M_{ln}d\rho_{kl}\right] \\ 
=-\sum_{kl}\left[\delta_{kl} +  (\ln\rho)_{kl}]\right]d\rho_{kl}\ ,
\end{multline*}
Q.E.D.

With the help of Lemma 1, Eq.\ (\ref{12.5}) becomes
$$
\text{Tr}\left[(1 + \ln\rho -\lambda_0-\lambda_1 q -\lambda_2 p-\lambda_3
  q^2-\lambda_4p^2)d\rho\right] = 0 
$$
so that we have
\begin{equation}\label{12.6}
\rho = \exp(-\lambda_0-1-\lambda_1 q-\lambda_2 p-\lambda_3 q^2-\lambda_4 p^2)\
. 
\end{equation}
The first two terms in the exponent determine the normalization constant
$$
e^{-\lambda_0-1}
$$
because they commute with the rest of the exponent and are independent of the
dynamical variables. Taking the trace of Eq.\ (\ref{12.6}), we obtain 
$$
e^{-\lambda_0-1} = \frac{1}{Z(\lambda_1,\lambda_2,\lambda_3,\lambda_4)}\ ,
$$
where $Z$ is the partition function,
\begin{equation}\label{13.1}
Z(\lambda_1,\lambda_2,\lambda_3,\lambda_4) = \text{Tr}[ \exp(-\lambda_1
q-\lambda_2 p-\lambda_3 q^2-\lambda_4 p^2)]\ . 
\end{equation}
Thus, the state operator has the form
\begin{equation}\label{13.2}
\rho = \frac{1}{Z(\lambda_1,\lambda_2,\lambda_3,\lambda_4)} \exp(-\lambda_1
q-\lambda_2 p-\lambda_3 q^2-\lambda_4 p^2)\ . 
\end{equation}

At this stage, the quantum theory begins to differ from the classical one. It
turns out that, for the case of non-commuting operators in the exponent of the
partition function, formula (\ref{43.1}) is not valid in general. We can only
show that it holds for the first derivatives. To this aim, we prove the
following 
\begin{lem}
Let $A$ and $B$ be hermitean matrices. Then
\begin{equation}\label{14.1}
\frac{d}{d\lambda}\text{Tr}[\exp(A+B\lambda)] = \text{Tr}[B\exp(A+B\lambda)]\ .
\end{equation}
\end{lem}
{\bf Proof}
We express the exponential function as a series and then use the invariance of
trace with respect to any cyclic permutation of its argument. 
\begin{multline*}
d\text{Tr}[\exp(A+B\lambda)] =
\sum_{n=0}^\infty\frac{1}{n!}\text{Tr}[d(A+B\lambda)^n] \\ 
= \sum_{n=0}^\infty\frac{1}{n!}\text{Tr}\left[\sum_{k=1}^n(A
  +B\lambda)^{k-1}B(A+B\lambda)^{n-k}\right]d\lambda \\ 
=\sum_{n=0}^\infty\frac{1}{n!}\sum_{k=1}^n\text{Tr}\left[B(A
  +B\lambda)^{n-1}\right]d\lambda = \text{Tr}[B\exp(A+B\lambda)]d\lambda\ , 
\end{multline*}
Q.E.D.
The proof of Lemma 2 shows why formula (\ref{43.1}) is not valid for higher
derivatives than the first in the quantum case: the operator $B$ does not
commute with $A+B\lambda$ and cannot be shifted from its position to the first
position in product  
$$
(A+B\lambda)^kB(A+B\lambda)^l\ .
$$
For the first derivative, it can be brought there by a suitable cyclic
permutation. However, each commutator $[B,(A+B\lambda)]$ is proportional to
$\hbar$. Hence, formula (\ref{43.1}) with higher derivatives  is the leading
term in the expansion of averages in powers of $\hbar$. 

Together with Eq.\ (\ref{13.1}), Lemma 2 implies the formulae:
\begin{equation}\label{13.3}
\frac{\partial \ln Z}{\partial\lambda_1} = -Q\ ,\quad \frac{\partial \ln
  Z}{\partial\lambda_3} = -Q^2-\Delta Q^2 
\end{equation}
and
\begin{equation}\label{13.4}
\frac{\partial \ln Z}{\partial\lambda_2} = -P\ ,\quad \frac{\partial \ln
  Z}{\partial\lambda_4} = -P^2-\Delta P^2\ . 
\end{equation}
The values of the multipliers can be calculated from Eqs.\ (\ref{13.3}) and
(\ref{13.4}), if the form of the partition function is known. 

Variational methods can find locally extremal values that are not necessarily
maxima. We can however prove that our state operator maximizes entropy. The
proof is based on the generalized Gibbs' inequality, 
$$
\text{Tr}(\rho\ln\rho-\rho\ln\sigma) \geq 0
$$
for all pairs $\{\rho,\sigma\}$ of state operators (for proof of the
inequality, see \cite{Peres}, P. 264). The proof of maximality is then
analogous to the 'classical' proof (see, e.g., \cite{Jaynes}, P. 357). The
first proof of maximality in the quantum case was given by von Neumann
\cite{JvN}. 

The state operator (\ref{13.2}) can be inserted in the formula (\ref{VNE}) to
give the value of the maximal entropy, 
\begin{equation}\label{15.1}
S = \ln Z + \lambda_1\langle q\rangle + \lambda_2\langle p\rangle +
\lambda_3\langle q^2\rangle + \lambda_4\langle p^2\rangle\ . 
\end{equation}
This, together with Eqs. (\ref{13.3}) and(\ref{13.4}) can be considered as the
Legendre transformation from the function $\ln
Z(\lambda_1,\lambda_2,\lambda_3,\lambda_4)$ to the function $S(\langle
q\rangle,\langle p\rangle,\langle q^2\rangle,\langle p^2\rangle )$.

\subsection{Diagonal representation}
The exponent in Eq. (\ref{13.2}) can be written in the form
\begin{equation}\label{19.1}
\frac{\lambda_1^2}{4\lambda_3} + \frac{\lambda_2^2}{4\lambda_4}
-2\sqrt{\lambda_3\lambda_4}K\ ,
\end{equation}
where
\begin{equation}\label{19.2}
K = \frac{1}{2}\sqrt{\frac{\lambda_3}{\lambda_4}}\left(q +
  \frac{\lambda_1}{2\lambda_3}\right)^2 +
\frac{1}{2}\sqrt{\frac{\lambda_4}{\lambda_3}}\left(p +
  \frac{\lambda_2}{2\lambda_4}\right)^2\ . 
\end{equation}
This is an operator acting on the Hilbert space of our system. $K$ has the
form of the Hamiltonian\footnote{The operator $K$ must not be confused with
  the Hamiltonian $H$ of our system, which can be arbitrary.} of a harmonic
oscillator with the coordinate $U$ and momentum $W$ 
\begin{equation}\label{16.1}
U = q + \frac{\lambda_1}{2\lambda_3}\ ,\quad W = p +
\frac{\lambda_2}{2\lambda_4}\ , 
\end{equation}
that satisfy the commutation relation $[U,W] = i\hbar$. The oscillator has
mass $M$ and frequency $\Omega$, 
\begin{equation}\label{16.2}
M = \sqrt{\frac{\lambda_3}{\lambda_4}}\ ,\quad \Omega = 1\ .
\end{equation}
The normalized eigenstates $|k\rangle$ of the operator form a basis in the
Hilbert space of our system defining the so-called {\em diagonal
  representation} and its eigenvalues are $\hbar/2 + \hbar k$. As usual, we
introduce operator $A$ such that  
\begin{eqnarray}\label{ua}
U &=& \sqrt{\frac{\hbar}{2M}}(A+A^\dagger)\ , \\
\label{qa}
W &=& -i\sqrt{\frac{\hbar M}{2}}(A-A^\dagger)\ , \\
\label{Hamilt}
K &=& \frac{\hbar}{2}(A^\dagger A + AA^\dagger))\ , \\
\label{aact1}
A|k\rangle &=& \sqrt{k}|k-1\rangle\ , \\
\label{aact2}
A^\dagger|k\rangle &=& \sqrt{k+1}|k+1\rangle\ .
\end{eqnarray}

To calculate $Z$ in the diagonal representation is easy:
\begin{multline*}
Z = \text{Tr}\left[\exp\left(\frac{\lambda_1^2}{4\lambda_3} +
    \frac{\lambda_2^2}{4\lambda_4} 
-2\sqrt{\lambda_3\lambda_4}K\right)\right] \\
= \sum_{k=0}^\infty\langle k|\exp\left(\frac{\lambda_1^2}{4\lambda_3} +
  \frac{\lambda_2^2}{4\lambda_4} 
-2\sqrt{\lambda_3\lambda_4}K\right)|k\rangle \\
= \exp\left(\frac{\lambda_1^2}{4\lambda_3} + \frac{\lambda_2^2}{4\lambda_4}
-\hbar\sqrt{\lambda_3\lambda_4}\right)
\sum_{k=0}^\infty\exp(-2\hbar\sqrt{\lambda_3\lambda_4}k)\ . 
\end{multline*}
Hence, the partition function for the quantum ME-packets is
\begin{equation}\label{17.3}
Z = \frac{\exp\left(\frac{\lambda_1^2}{4\lambda_3} +
    \frac{\lambda_2^2}{4\lambda_4}\right)}
{2\sinh(\hbar\sqrt{\lambda_3\lambda_4})}\ .  
\end{equation}

Now, we can express the Lagrange multipliers in terms of the averages and
variances. Eqs.\ (\ref{13.3}) and (\ref{13.4}) yield 
\begin{equation}\label{18.1}
\lambda_1 = -\frac{Q}{\Delta Q^2}\frac{\nu}{2}\ln\frac{\nu+1}{\nu-1}\ ,\quad
\lambda_2 = -\frac{P}{\Delta P^2}\frac{\nu}{2}\ln\frac{\nu+1}{\nu-1}\ ,
\end{equation}
and
\begin{equation}\label{18.2}
\lambda_3 = \frac{1}{2\Delta Q^2}\frac{\nu}{2}\ln\frac{\nu+1}{\nu-1}\ ,\quad
\lambda_4 = \frac{1}{2\Delta P^2}\frac{\nu}{2}\ln\frac{\nu+1}{\nu-1}\ , 
\end{equation}
where $\nu$ is defined by Eq.\ (\ref{uncertainty})

From Eq.\ (\ref{15.1}), (\ref{18.1}) and (\ref{18.2}), we obtain the entropy:
\begin{equation}\label{18.5}
S = -\ln 2 + \frac{\nu+1}{2}\ln(\nu+1) -\frac{\nu-1}{2}\ln(\nu-1)\ .
\end{equation}
Thus, $S$ depends on $Q$, $P$, $\Delta Q$, $\Delta P$ only via $\nu$. We have 
$$
\frac{dS}{d\nu} = \frac{1}{2}\ln\frac{\nu+1}{\nu-1} > 0\ ,
$$
so that $S$ is an increasing function of $\nu$. Near $\nu = 1$,
$$
S \approx -\frac{\nu-1}{2}\ln(\nu-1)\ .
$$
Asymptotically ($\nu \rightarrow\infty$),
$$
S \approx \ln\nu+1-\ln 2\ .
$$
In the classical region, $\nu \gg 1$, $S\approx\ln\nu$.

It is clear that the choice of Q and P cannot influence the entropy. The
independence of $S$ from $Q$ and $P$ does not contradict the Legendre
transformation properties. Indeed, usually, one would have 
$$
\frac{\partial S}{\partial Q} = \lambda_1\ ,
$$
but here
$$
\frac{\partial S}{\partial Q} = \lambda_1 + 2\lambda_3 Q\ ,
$$
which is zero.

The state operator can also be expressed in terms of the averages and
variances. The trivial generalization to $n$ degrees of freedom is 
\begin{thm}
The state operator of the ME-packet of a system with given averages and
variances $Q_1,\cdots,Q_n$, $\Delta Q_1,\cdots,\Delta Q_n$ of coordinates and
$P_1,\cdots,P_n$, $\Delta P_1,\cdots,\\ \Delta P_n$ of momenta, is 
\begin{equation}\label{32.1}
\rho =
\prod_{k=1}^n\left[\frac{2}{\nu_k^2-1}\exp\left(
    -\frac{1}{\hbar}\ln\frac{\nu_k+1}{\nu_k-1}K_k\right)\right]\ , 
\end{equation}
where
\begin{equation}\label{20.3}
K_k = \frac{1}{2}\frac{\Delta P_k}{\Delta Q_k}(q_k-Q_k)^2 +
\frac{1}{2}\frac{\Delta Q_k}{\Delta P_k}(p_k-P_k)^2 
\end{equation}
and 
\begin{equation}\label{20.3b}
\nu_k = \frac{2\Delta P_k\Delta Q_k}{\hbar}\ .
\end{equation}
\end{thm}
Strictly speaking, the state operator (\ref{32.1}) is not a Gaussian
distribution. Thus, it  seems to be either a counterexample to, or a
generalization of, Jaynes statement that the Gaussian distribution is the only
distribution that maximizes entropy for given values of the first two moments
\cite{Jaynes}. 

In the diagonal representation, we have
\begin{equation}\label{20.2a}
\rho = \sum_{k=0}^\infty R_k|k\rangle\langle k|\ .
\end{equation}
We easily obtain for $R_k$ that
\begin{equation}\label{20.2}
R_k = 2\frac{(\nu-1)^k}{(\nu+1)^{k+1}} .
\end{equation}
Hence,
$$
\lim_{\nu=1}R_k = \delta_{k0}\ ,
$$
and the state $\rho$ becomes $|0\rangle\langle 0|$. In general, states
$|k\rangle$ depend on $\nu$. The state vector $|0\rangle$ in the
$q$-representation expressed as a function of $Q$, $P$, $\Delta Q$ and $\nu$
is given by 
\begin{equation}\label{20.1}
\psi(q) = \left(\frac{1}{\pi} \frac{\nu}{2\Delta Q^2}\right)^{1/4}
\exp\left[-\frac{\nu}{4\Delta Q^2}(q-Q)^2 + \frac{iPq}{\hbar}\right]\ . 
\end{equation}
This is a Gaussian wave packet that corresponds to different values of
variances than the original ME packet but has the minimal uncertainty. For
$\nu\rightarrow 1$, it remains regular and the projector $|0\rangle\langle 0|$
becomes the state operator of the original ME packet. Hence, Gaussian wave
packets are special cases of ME-packets. 

The diagonal representation offers a method for calculating averages of
coordinates and momenta products that replaces the partition function way. Let
us denote such a product X. We have 
\begin{equation}\label{avX}
\langle X\rangle = \sum_{k=0}^\infty R_k\langle k|X|k\rangle\ .
\end{equation}
To calculate $\langle k|X|k\rangle$, we use Eqs.\ (\ref{ua}), (\ref{qa}),
(\ref{16.1}), (\ref{16.2}), (\ref{18.1}) and (\ref{18.2}) to obtain 
$$
q = Q +\frac{\Delta Q}{\sqrt{\nu}}(A+A^\dagger)\ ,\quad p = P-i\frac{\Delta
  P}{\sqrt{\nu}} (A-A^\dagger)\ . 
$$
By substituting these relations to $X$ and using the commutation relations
$[A,A^\dagger] = 1$, we obtain  
$$
X = {\mathcal P}(N) + {\mathcal Q}(A,A^\dagger)\ ,
$$
where $N = A^\dagger A$ and where, in each monomial of the polynomial
${\mathcal Q}$, the number of  $A$-factors is different from the number of
$A^{\dagger}$-factors. Thus, 
$$
\langle k|X|k\rangle = {\mathcal P}(k)\ .
$$
In Eq.\ (\ref{avX}), there are, therefore, sums
$$
\sum_{k=0}^\infty k^nR_k\ .
$$
With Eq.\ (\ref{20.2}), this becomes
$$
\sum_{k=0}^\infty k^nR_k = \frac{2}{\nu+1}I_n\ ,
$$
where
$$
I_n(\nu) = \sum_{k=0}^\infty k^n\left(\frac{\nu-1}{\nu+1}\right)^k\ .
$$
We easily obtain
$$
I_n = \left(\frac{\nu^2-1}{2}\frac{d}{d\nu}\right)^n\frac{\nu+1}{2}\ .
$$
The desired average value is then given by
\begin{equation}\label{average}
\langle X \rangle = \frac{2}{\nu+1}{\mathcal
  P}\left(\frac{\nu^2-1}{2}\frac{d}{d\nu}\right)\frac{\nu+1}{2}\ . 
\end{equation}
The calculation of the polynomial ${\mathcal P}$ for a given $X$ and the
evaluation of the right-hand side of Eq.\ (\ref{average}) are the two steps of
the promised method.

\subsection{Equations of motion}
Let the Hamiltonian of our system be $H$ and the unitary evolution group
$U(t)$. The dynamics in the Schr\"{o}dinger picture leads to the time
dependence of $\rho$: 
$$
\rho(t) = U(t)\rho U^\dagger(t)\ .
$$
Substituting for $\rho$ from Eq.\ (\ref{32.1}) and using a well-known property
of exponential function, we obtain 
\begin{equation}\label{32.2}
\rho(t) = \frac{2}{\nu^2-1}\exp\left(-\frac{1}{\hbar}\ln\frac{\nu+1}{\nu
    -1}U(t)KU^\dagger(t)\right)\ . 
\end{equation}

In the Heisenberg picture, $\rho$ remains constant, while $q$ and $p$ are time
dependent and satisfy the equations 
\begin{equation}\label{33.1}
i\hbar\frac{dq}{dt} = [q,H]\ ,\quad i\hbar\frac{dp}{dt} = [p,H]\ .
\end{equation}
They are solved by
$$
q(t) = U^\dagger(t)qU(t)\ ,\quad p(t) = U^\dagger(t)pU(t)\ ,
$$
where $q$ and $p$ are the initial operators, $q=q(0)$ and $p=p(0)$. The
resulting operators can be written in the form of operator functions 
analogous to classical expressions (\ref{36.65}) so that Eqs.\ (\ref{36.7})
and (\ref{36.8}) can again be used. 

The example with potential function (\ref{36.1}) is solvable in quantum
theory, too, and we can use it for comparison with the classical dynamics as
well as for a better understanding of the ME-packet dynamics. Eqs.\
(\ref{33.1}) have then the solutions given by (\ref{37.1}) and (\ref{37.2})
with functions $f_n(t)$ and $g_n(t)$ given by (\ref{37.4}) and (\ref{37.5}) or
(\ref{37.9a}) and (\ref{37.9b}). The calculation of the averages and variances
is analogous to the classical one and we obtain  Eqs.\ (\ref{38.3}) and
(\ref{38.4}) again with the difference that the term $2\langle qp\rangle$ on
the right hand side of (\ref{38.4}) is now replaced by $\langle
qp+pq\rangle$. 

To calculate $\langle qp+pq\rangle$, we use the method introduced in the
previous section. We have 
$$
qp+pq = 2QP + 2\frac{P\Delta Q}{\sqrt{\nu}}(A+A^\dagger)- 2i\frac{Q\Delta
  P}{\sqrt{\nu}}(A-A^\dagger)-2i\frac{\Delta Q\Delta P}{\nu} (A^2-A^{\dagger
  2})\ . 
$$
hence, ${\mathcal P} = 2QP$, and 
$$
\langle qp+pq\rangle = 2QP\ .
$$
The result is again Eq.\ (\ref{39.1}). Similarly for $p$, the results are
given by Eqs.\ (\ref{39.2}) and (\ref{39.3}).  

We have shown that the averages and variances of quantum ME-packets have
exactly the same time evolution as those of classical ME-packets in the
special case of at-most-quadratic potentials. From formulae (\ref{39.1}) and
(\ref{39.3}) we can also see an interesting fact. On the one hand, both
variances must increase near $t=0$. On the other, the entropy must stay
constant because the evolution of the quantum state is unitary. As the
relation between entropy and $\nu$ is fixed for ME-packets, the ME-packet form
is not preserved by the evolution (the entropy ceases to be maximal). This is
similar for Gaussian-packet form or for coherent-state form. 

For general potentials, there will be two types of corrections to the dynamics
of the averages: terms containing the variances and terms containing
$\hbar$. To see these corrections, let us calculate time derivatives for the
Hamiltonian (\ref{35.1}) with potential (\ref{50.2}). The Heisenberg-picture
equations of motion give again 
$$
\frac{dq}{dt} = \frac{1}{m}p\ ,
$$
so that Eq.\ (\ref{50.4}) is valid. The other equation,
$$
i\hbar\frac{dp}{dt} = [p,H]\ ,
$$
can be applied iteratively as in the classical case so that all time
derivatives of $p$ can be obtained. Thus, 
\begin{equation}\label{52.1}
\frac{dp}{dt} = -V_1-V_2q-\frac{V_3}{2}q^2-\frac{V_4}{6}q^3 + r_5\ ,
\end{equation}
and
$$
\frac{d^2p}{dt^2} = -\frac{V_2}{m}p - \frac{V_3}{2m}(qp+pq) -
\frac{V_4}{6m}(q^2p+qpq+pq^2) + r_5\ . 
$$
This differs from the classical equation only by factor ordering. We can use
the commutator $[q,p]=i\hbar$ to simplify the last term, 
\begin{equation}\label{52.2}
\frac{d^2p}{dt^2} = -\frac{V_2}{m}p-\frac{V_3}{2m}(qp+pq)-\frac{V_4}{2m}qpq +
r_5\ . 
\end{equation}
Similarly,
\begin{multline}\label{52.3}
\frac{d^3p}{dt^3} = -\frac{V_3}{m^2}p^2-\frac{V_4}{m^2}pqp + \frac{V_1V_2}{m}
+ \frac{V_1V_3+V_2^2}{m}q + \frac{3V_2V_3+V_1V_4}{2m}q^2 \\ +
\frac{4V_2V_4+3V_3^2}{6m}q^3 + \frac{5V_3V_4}{12m}q^4 + \frac{V_4^2}{12m}q^5 +
r_5\ , 
\end{multline}
and
\begin{multline}\label{52.4}
\frac{d^4p}{dt^4} = -\frac{V_4}{m^3}p^3 + \frac{3V_1V_3+V_2^2}{m^2}p +
\frac{3V_1V_4+5V_2V_3}{2m^2}(qp+pq) + \frac{5V_3^2+8V_2V_4}{2m^2}qpq \\ +
\frac{3V_3V_4}{2m^2}(q^3p+pq^3) + \frac{3V_4^2}{4m^2}q^2pq^2 + r_5\ . 
\end{multline}

Next, we calculate quantum averages with the help of formula
(\ref{average}). The quantum averages of the monomials that are linear in one
of variables $q$ or $p$ can differ from their classical counterparts only by
terms that are of the first order in $1/\nu$ and purely imaginary. For
example, 
$$
\langle qp\rangle = QP + \frac{i\hbar}{2}\ ,
$$
or
$$
\langle q^3p\rangle = Q^3P + 3QP\Delta Q^2 + 3i\frac{Q^2\Delta Q\Delta P}{\nu}
+ 3i\frac{\Delta Q^3\Delta P}{\nu}\ . 
$$
These corrections clearly cancel for all symmetric factor orderings. The first
term in which a second-order correction occurs is $q^2p^2$ and we obtain for
it: 
$$
\langle pq^2p\rangle = \langle q^2p^2\rangle_{\text{class}} + 2\frac{\Delta
  Q^2\Delta P^2}{\nu^2}\ . 
$$

The equations (\ref{52.1})-(\ref{52.4}) do not contain any such terms and so
their averages coincide exactly with the classical equations
(\ref{51.1})-(\ref{51.4}). The terms $q^2p^2$ with different factor orderings
occur in the fifth time derivative of $p$ and have the form 
$$
\frac{3V_3V_4}{2m^2}\left[q^3p+pq^3,\frac{p^2}{2m}\right] +
\frac{V_3V_4}{2m^3}\left[\frac{1}{3}q^3,p^3\right]  
=i\hbar\frac{V_3V_4}{2m^3}(21pq^2p-11\hbar^2)\ .
$$
The average of the resulting term in the fifth time derivative of $p$ is
$$
\frac{V_3V_4}{2m^3}\left(21Q^2P^2 + 21P^2 \Delta Q^2 + 21 Q^2\Delta P^2 +
  21\Delta Q^2\Delta P^2-\frac{\hbar^2}{2}\right)\ . 
$$
If we express $\hbar$ as $2\Delta Q\Delta P/\nu$, we can write the last two
terms in the parentheses as 
$$
\Delta Q^2\Delta P^2\left(21-\frac{2}{\nu^2}\right)\ .
$$
A similar term appears in the third time derivative of $p$, if we allow
$V_5\neq 0$ in the expansion (\ref{50.2}): 
$$
\left[-\frac{V_5}{12m}(q^3p+pq^3),\frac{p^2}{2m}\right] =
i\hbar\left[-\frac{V_5}{4m^2}(2pq^2p+\hbar^2)\right]\ , 
$$
which contributes to $d^3P/dt^3$ by
$$
-\frac{V_5}{2m^2}\left(\langle q^2p^2\rangle_{\text{class}} + \frac{4\Delta
    Q^2\Delta P^2}{\nu^2}\right)\ . 
$$
Again, the correction is of the second order in $\nu^{-1}$.

We can conclude. The quantum equations begin to differ from the classical
one's only in the higher order terms in $V$ or in the higher time derivatives
and the correction is of the second order in $1/\nu$. This seems to be very
satisfactory: our quantum model reproduces the classical dynamic very
well. Moreover, Eq.\ (\ref{20.1}) shows that Gaussian wave packets are special
cases of ME packets with $\nu=1$. Thus, they approximate classical
trajectories less accurately than ME packets with large $\nu$.

\section{Classical limit}
At some places of the paper, it is written that $\nu\gg 1$ is the classical
regime. Let us now look to see if our equations give some support to this
statement. 

Let us consider averages of powers of $q$ and $p$. If we expand such an
average in powers of Planck constant then the leading term can be calculated
with help of the formula (\ref{43.1}) from quantum partition function
(\ref{17.3}) and from relations (\ref{18.1}) and (\ref{18.2}) between the
Lagrange multipliers and the averages and variances of $q$ and $p$. This has
been explained after the proof of Lemma 2. 

The quantum partition function (\ref{17.3}) differs from its classical
counterpart (\ref{22.1}) by the denominator
$\sinh(\hbar\sqrt{\lambda_3\lambda_4})$. If  
\begin{equation}\label{ll} 
\hbar\sqrt{\lambda_3\lambda_4} \ll 1\ ,
\end{equation}
we can write
$$
\sinh(\hbar\sqrt{\lambda_3\lambda_4}) = \hbar\sqrt{\lambda_3\lambda_4}[1 +
O((\hbar\sqrt{\lambda_3\lambda_4})^2)] 
$$
The leading term in the partition function then is
$$
Z = \frac{\pi}{h}\frac{1}{\sqrt{\lambda_3\lambda_4}}
\exp\left(\frac{\lambda_1^2}{4\lambda_3} +
  \frac{\lambda_2^2}{4\lambda_4}\right)\ , 
$$
where $h = 2\pi\hbar$. Comparing this with formula (\ref{22.1}) shows that the
two expressions are identical, if we set 
$$
v = h\ .
$$
We can say that quantum mechanics gives us the value of $v$. Next, we have to
express condition (\ref{ll}) in terms of the averages and variances. Equations
(\ref{18.2}) imply 
$$
\hbar\sqrt{\lambda_3\lambda_4} = \frac{1}{2}\ln\frac{\nu+1}{\nu-1}\ .
$$
Hence, condition (\ref{ll}) is equivalent to
\begin{equation}\label{gg}
\nu \gg 1\ .
\end{equation}
The expression
$$
\frac{\nu}{2}\ln\frac{\nu+1}{\nu-1}
$$
that appears on the right-hand sides of Eqs.\ (\ref{18.1}) and (\ref{18.2})
satisfies 
$$
\lim_{\nu=\infty}\frac{\nu}{2}\ln\frac{\nu+1}{\nu-1} = 1\ .
$$
Hence, the leading terms in these equations coincide with Eqs.\ (\ref{22.2})
and (\ref{22.3}). 

Our result can be formulated as follows. The time evolution of classical and
quantum ME packets with the same initial values of averages and variances
defines the averages as time functions. These time functions coincide for the
two theories in the limit $\nu = \infty$. Hence, in our approach, this is the
classical limit. It is very different from the usual assumption that the
classical limit must yield the variances as small as possible. One also often
requires that commutators of observables vanish in classical limit. This is
however only motivated by the assumption that all basic quantum properties are
single values of observables. Within our interpretation, this assumption is
rejected and if classical observables are related to quantum operators then
only by being average values of the operators in prepared states. All such
averages are defined by the preparation and do exist simultaneously,
independently of whether the operators commute or not. For example, $Q$ and
$P$ are such simultaneously existing variables for ME packets. 

Let us compare the present paper notion of classical limit with a modern
textbook version such as Ch.\ 14 of \cite{Ballentine}. Both approaches define
the classical limit of a quantum state as a classical ensemble described by a
fuzzy distribution function and calculate time evolutions of averages in the
states. However, in the textbook, any quantum system, even not macroscopic,
and any state, even pure, are allowed (pure states preferred as they have
smaller uncertainties). Hence, our notion is much narrower: we consider only
macroscopic quantum systems and only some of their maximum entropy
states. This has obvious physical reasons explained in the first two
sections.

\section{Conclusion}
The paper describes a quite general construction of quantum states that model
important properties of classical-mechanical states. To achieve that, one
often-assumed classical property has to be abandoned: the completely sharp
trajectory of all mechanical systems. The sharp trajectory is considered here
only as an ideal limit allowed by classical mechanics. There is however
nothing in nature that corresponds to it. This is in agreement both with
practical observations and with theoretical idea that the correct underlying
theory is quantum mechanics. Hence, the way for statistical methods
highlighted in Ref. \cite{PHJT} is free. The key concept turned out to be
physical conditions equivalent to preparation process in quantum
mechanics. The paper transfers it into classical mechanics, where it
generalises the old notion of initial data. Entropy is defined by the physical
conditions independently of the state of any observer's mind. 

Classical mechanics allows not only sharp, but also fuzzy trajectories and the
comparison of some classical and quantum fuzzy trajectories shows a very good
match. The fuzzy states chosen here are the so-called ME packets. Their
fuzziness is described by the quantity $\nu = 2\Delta Q\Delta P/\hbar$. The
entropy of an ME packet depends only on $\nu$ and is an increasing function of
it. The larger $\nu$ is, the better the quantum and the classical  evolutions
of average values have been shown to agree. Thus, the classical regime is
neither $\Delta Q = \Delta P = 0$ (absolutely sharp trajectory) nor $\nu=1$
(minimum quantum uncertainty). This is the most important result of the
paper. 

Unlike internal classical properties the external ones such as coordinates and
momenta are well manipulable so that conditions exist allowing ME packets to
have all variances $\Delta Q$ and $\Delta P$ from a broad range. There is no
reason why the variances {\em had always to be} small other than the incorrect
assumption that all real mechanical trajectories are absolutely sharp. Hence,
our account of classical properties as statistical properties of macroscopic
quantum systems that started in \cite{PHJT} can be considered as concluded.

\subsection*{Acknowledgements}
The author is indebted to Ferenc Niedermayer and Jiri Tolar for discussions.

\end{document}